\documentclass[footinbib,aps,pra,reprint,superscriptaddress,nopacs]{revtex4-1}
\usepackage{amsmath,amssymb,bbm,bm,graphicx}
\usepackage[absolute]{textpos}
\usepackage{graphicx}
\usepackage{xcolor}

\DeclareMathOperator{\Tr}{Tr}

\begin{document}
\title{A controlled-NOT gate for frequency-bin qubits}

\author{Hsuan-Hao Lu}
\thanks{These authors contributed equally to this work.}
\affiliation{School of Electrical and Computer Engineering and Purdue Quantum Center, Purdue University, West Lafayette, Indiana 47907, USA}
\author{Joseph M. Lukens}
\thanks{These authors contributed equally to this work.}
\affiliation{Quantum Information Science Group, Computational Sciences and Engineering Division, Oak Ridge National Laboratory, Oak Ridge, Tennessee 37831, USA}
\author{Brian P. Williams}
\affiliation{Quantum Information Science Group, Computational Sciences and Engineering Division, Oak Ridge National Laboratory, Oak Ridge, Tennessee 37831, USA}
\author{Poolad Imany}
\affiliation{School of Electrical and Computer Engineering and Purdue Quantum Center, Purdue University, West Lafayette, Indiana 47907, USA}
\author{Nicholas A. Peters}
\affiliation{Quantum Information Science Group, Computational Sciences and Engineering Division, Oak Ridge National Laboratory, Oak Ridge, Tennessee 37831, USA}
\affiliation{Bredesen Center for Interdisciplinary Research and Graduate Education, The University of Tennessee, Knoxville, Tennessee 37996, USA}
\author{Andrew M. Weiner}
\affiliation{School of Electrical and Computer Engineering and Purdue Quantum Center, Purdue University, West Lafayette, Indiana 47907, USA}
\author{Pavel Lougovski}
\email{lougovskip@ornl.gov}
\affiliation{Quantum Information Science Group, Computational Sciences and Engineering Division, Oak Ridge National Laboratory, Oak Ridge, Tennessee 37831, USA}
\date{\today}


\begin{abstract}
The realization of strong photon-photon interactions has presented an enduring challenge across photonics, particularly in quantum computing, where two-photon gates form essential components for scalable quantum information processing (QIP)~\cite{Nielsen2000}. While linear-optic schemes have enabled probabilistic entangling gates in spatio-polarization encoding~\cite{Knill2001, Kok2007}, solutions for many other useful degrees of freedom remain missing. In particular, no two-photon gate for the important platform of frequency encoding~\cite{Kobayashi2016, Clemmen2016, Lu2018a, Lu2018b} has been experimentally demonstrated, due in large part to the additional challenges imparted by the mismatched wavelengths of the interacting photons. In this article, we design and implement the first entangling gate for frequency-bin qubits, a coincidence-basis controlled-NOT (CNOT), using line-by-line pulse shaping and electro-optic modulation. We extract a quantum gate fidelity of $0.91 \pm 0.01$ via a novel parameter inference approach based on Bayesian machine learning, which enables accurate gate reconstruction from measurements in the two-photon computational basis alone. Our CNOT imparts a single-photon frequency shift controlled by the frequency of another photon---an important capability in itself---and should enable new directions in fiber-compatible QIP.
\end{abstract}

\maketitle

As carriers of quantum information, optical photons feature a host of valuable attributes, such as immunity to environmentally induced decoherence, availability of precise tools for state control, and room temperature operation,  enabling quantum information processing (QIP) in a variety of encodings such as space/polarization~\cite{Kok2007,Babazadeh2017} and temporal modes~\cite{Humphreys2013,Ansari2018}.
Frequency-bin encoding---which offers additional advantages in terms of compatibility with state-of-the-art fiber-optic networks---has advanced rapidly in recent years, facilitated by the development of integrated frequency-bin photon sources~\cite{Reimer2016, Jaramillo2017, Kues2017, Imany2018} and quantum gates based on both nonlinear-optical~\cite{Kobayashi2016, Clemmen2016} and electro-optical~\cite{Lu2018a, Lu2018b} mixing approaches. However,  two-photon entangling gates for frequency bins have yet to be realized on any platform. 

Such entangling gates are required for universal QIP, for an arbitrary quantum operation can be constructed with single-qubit rotations plus a two-qubit entangling gate~\cite{Nielsen2000}. While photonics excels for single-qubit gates, the inherent difficulty in realizing photon-photon interactions has made the two-qubit gate a persistent obstacle in photonic QIP. In the absence of a sufficient nonlinearity, such gates can still be achieved via quantum interference, ancilla photons, and single-photon detection. While two-qubit gates succeed only probabilistically in this paradigm, linear-optical quantum computation (LOQC)~\cite{Knill2001} is in principle scalable with polynomial auxiliary resource requirements and has laid the foundation for many subsequent advances in photonic QIP~\cite{Pittman2001,Pittman2002,Ralph2002,Hofmann2002,OBrien2003,Kok2007}. It is this approach which we invoked in proposing \emph{spectral} LOQC---a universal QIP scheme tailored to frequency-bin qubits which makes use of electro-optic phase modulators (EOMs) and Fourier-transform pulse shapers (PSs)~\cite{Lukens2017}. Spectral LOQC has been utilized to demonstrate coherent single-photon operations with near-unity fidelity~\cite{Lu2018a, Lu2018b}, but a two-photon gate has heretofore proven elusive.



Theoretically, we previously discovered EOM/PS configurations capable of realizing ancilla-based two-qubit gates in spectral LOQC~\cite{Lukens2017}. Yet if one relaxes the gate requirements slightly, by conditioning on the presence of a photon in each pair of qubit modes, it is well-known in standard LOQC that one can engineer a two-qubit gate with no ancillas and success probability $\mathcal{P}=1/9$~\cite{Ralph2002, Hofmann2002}. Assuming a quantum nondemolition measurement is unavailable, such gates are destructive (succeeding only when both information-carrying photons are detected). Yet they require only two-fold coincidences for characterization, making them excellent choices for experimental studies of basic quantum computing functionalities.


To explore two-qubit coincidence-basis gates in spectral LOQC, we follow the optimization approach in Refs. \cite{Lukens2017, Lu2018a}, numerically finding phase patterns for an EOM/PS sequence which maximize success $\mathcal{P}$ constrained to fidelity $\mathcal{F}\geq0.9999$. Specifically, with $U_\mathrm{ideal}$ defined as the desired two-qubit unitary and $W$ the actual Hilbert space transformation,
\begin{equation}
\label{e1}
\mathcal{P} = \frac{\Tr W^\dagger W}{d}
\end{equation}
\begin{equation}
\label{e2}
\mathcal{F} = \frac{\left| \Tr (U_\mathrm{ideal}^\dagger W) \right|^2}{d^2 \mathcal{P}},
\end{equation}
where $d=4$ is the dimensionality of the subspace spanned by the coincidence basis~\cite{Lukens2017}. In order to facilitate experimental implementation, we restrict our simulations to sinewave-only electro-optic modulation. We find that a 3EOM/2PS sequence can realize a frequency-bin CNOT at the optimal success probability of $\mathcal{P}=1/9$, while a smaller 2EOM/1PS circuit can do so with reduced success: $\mathcal{P}=0.0445$. (See Fig.~\ref{figE1} for the specific EOM/PS modulation patterns.) Due to equipment availability and system complexity, we elect to implement this simpler 2EOM/1PS CNOT in the experiments below.



\begin{figure}
\centering
\includegraphics[width=\columnwidth]{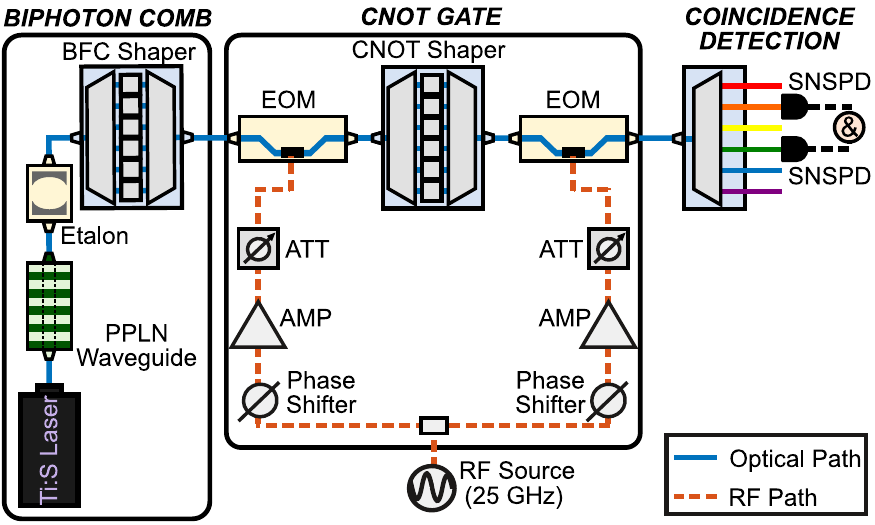}
\caption{Experimental setup (see text for details). PPLN: periodically poled lithium niobate, EOM: electro-optic phase modulator, ATT: variable radio-frequency (RF) attenuator, AMP: RF amplifier, SNSPD: superconducting nanowire single-photon detecctor.}
\label{fig1}
\end{figure}

\begin{figure*}[!t]
\centering
\includegraphics[width=\textwidth]{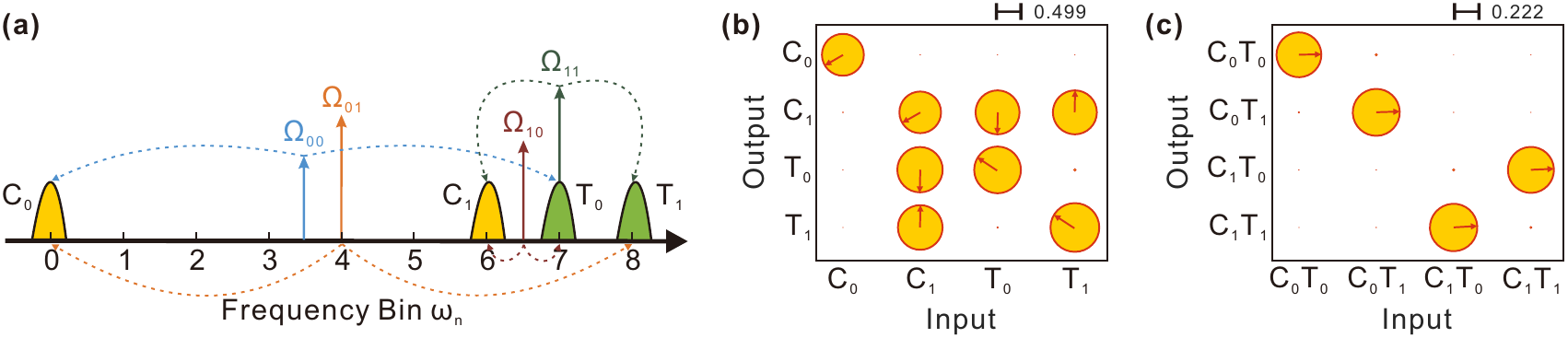}
\caption{(a) Mode definitions for frequency-bin control and target qubits. The labels $\{\Omega_{00},\Omega_{01},\Omega_{10},\Omega_{11}\}$ mark the pump frequency values (divided by two) needed to produce each of the computational basis states. (b) Experimentally obtained complex mode transformation $V$. (c) Inferred two-photon transformation $W$ obtained from permanents of $2\times 2$ submatrices of $V$.}
\label{fig2}
\end{figure*}

Figure~\ref{fig1} provides a schematic of the setup. The gate itself comprises the central EOM/PS/EOM sequence, and the frequency bins for encoding are defined according to $\omega_n = \omega_0 + n\Delta\omega$, where $\omega_0 = 2\pi \times 193.45$ THz and $\Delta\omega = 2\pi \times 25$ GHz, corresponding to the standard ITU grid and facilitating low-crosstalk, line-by-line shaping by our 10 GHz resolution pulse shapers. The specific bins for encoding follow in Fig.~\ref{fig2}(a), where $\{C_0,C_1\}$ and $\{T_0,T_1\}$ denote logical $|0\rangle$ and $|1\rangle$ for the control and target, respectively. This particular mode placement makes sense conceptually: mode $C_0$ is spectrally isolated from the target's logical bins, ensuring a photon in mode $C_0$ leaves the target unchanged; on the other hand, bin $C_1$ is close to both target bins, able to be coupled to $T_0$ and $T_1$ with equal strength.

Since this gate is based on a linear-optical network, we can estimate its performance using coherent-state-based characterization~\cite{Lu2018a, Rahimi2013}, i.e., probing it with an electro-optic frequency comb and measuring the output spectrum for different input frequency superpositions. This technique allows us to estimate the mode transformation matrix $V$, which controls how input mode operators $\hat{a}_n$ at each frequency $\omega_n$ transform to the output operators $\hat{b}_n$: $\hat{b}_n = \sum_{n^\prime} V_{nn^\prime} \hat{a}_{n^\prime}$. The mode matrix $V$, averaged over five independent measurements and projected onto the four computational modes, is shown in Fig.~\ref{fig2}(b). We use phasor notation to represent the complex elements $V_{nn^\prime}$; the radius signifies the amplitude, with the scale bar showing the maximum value in the matrix (0.499) and the arrow marking out the phase. (See Appendix B for values of all matrix elements including uncertainty.) From this matrix $V$, we can compute the equivalent two-photon state transformation matrix $W$~\cite{Lukens2017}, plotted for the coincidence basis in Fig.~\ref{fig2}(c) and also normalized to its peak magnitude of 0.222.

Because this estimate predicts all four of the large elements of $W$ to be in-phase, the corresponding inferred fidelity is $\mathcal{F}_\mathrm{inf} =0.995\pm0.001$; the success probability is $\mathcal{P}_\mathrm{inf} =0.0460\pm0.0005$. Both values are with respect to the ideal CNOT and in good agreement with theory. We emphasize that, unlike single-qubit gates which act on photons independently, two-qubit entangling gates rely on quantum interference effects that are inherently absent with high-flux laser fields. Thus this inferred fidelity is only an indirect estimate, based on extrapolating measured one-photon interference results to the two-photon case. Nevertheless, it provides strong initial evidence for the phase coherence and proper operation of our gate.

To test our gate with truly quantum states, however, we prepare a biphoton frequency comb (BFC) by pumping a periodically poled lithium niobate (PPLN) waveguide with a continuous-wave Ti:sapphire laser under type-0 phase matching, followed by filtering with a Fabry-Perot etalon with 25 GHz mode spacing and a full-width at half-maximum linewidth of 1.8 GHz (see Fig.~\ref{fig1}). The BFC pulse shaper subsequently selects specific modes as input to the gate. By translating the pump frequency to four different values [as shown in Fig.~\ref{fig2}(a)] and filtering out all but the desired modes using the BFC pulse shaper, we can prepare all inputs from the two-qubit computational basis: $|C_0  T_0\rangle = |1_{\omega_0} 1_{\omega_7}\rangle$, $|C_0  T_1\rangle = |1_{\omega_0} 1_{\omega_8}\rangle$, $|C_1  T_0\rangle = |1_{\omega_6} 1_{\omega_7}\rangle$, and $|C_1  T_1\rangle = |1_{\omega_6} 1_{\omega_8}\rangle$. To ensure the photon flux remains constant across the four inputs, we tune the PPLN waveguide temperature to align the peak of the phase-matching spectrum with the pump laser frequency. After the gate, the output photons are frequency-demultiplexed: we send control photon bins to detector $A$ and target photon bins to detector $B$.

\begin{figure}[!b]
\centering
\includegraphics[width=\columnwidth]{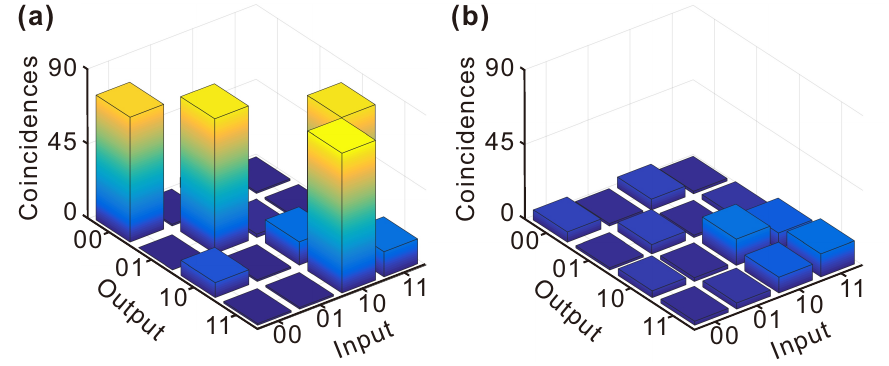}
\caption{(a) Coincidences registered over 600 s for all input/output state combinations. (b) Estimated number of accidentals computed from the product of single detector counts.}
\label{fig3}
\end{figure}

Figure~\ref{fig3}(a) shows the measured coincidences for all 16 input/output mode combinations, integrated over 600~s for each point. As expected, inputs with a photon in control mode 0 retain their quantum state, whereas a photon in control mode 1 leads to a flip in the output target qubit. Incidentally, the fraction of raw counts registered in the correct output state, averaged over all computational-basis inputs, is 87\%, comparable to the 84\% observed in the first coincidence-basis CNOT, which utilized spatio-polarization encoding~\cite{OBrien2003}. In Fig.~\ref{fig3}(b) we plot the accidentals as determined by the product of the singles counts and our timing resolution~\cite{Eckart1938}. The nonuniform distribution of accidentals stems from the fact that the singles counts vary significantly across input/output state combinations. Indeed, this is a natural feature of coincidence-basis gates: they are designed to discard cases when one of the qubit spaces is empty or doubly occupied, so that photon detection rates in a specific mode can change without impacting the designed operation.


Such information-bearing features in the accidentals suggest that incorporating knowledge from single-detector events---as well as the coincidences---can add significant value for  quantifying the performance of our gate in the presence of noise. To utilize all of our experimental data in a consistent fashion, we make use of Bayesian machine learning techniques to implement a numerical parameter inference approach built on Bayesian mean estimation (BME)~\cite{MacKay2003}. In the context of quantum state retrieval, BME is a powerful method which returns uncertainties on any quantity directly and makes efficient use of all available information, in the sense that the confidence in any estimate naturally reflects the amount of data gathered~\cite{Blume2010}. BME models for photon pairs including single-detector events have been developed as well, permitting extraction of the quantum pathway efficiencies in conjunction with estimates of the input density matrix~\cite{Williams2017}. In our BME model here, not only do we account for noise effects, but we can also retrieve meaningful estimates of the full complex matrix $V$, even though we only prepare and measure states in the computational basis. This represents an entirely new capability in two-photon gate analysis, for previously such truth-table measurements as in Fig.~\ref{fig3}(a) have only been used to establish \emph{magnitudes} in the matrix transformation, with superposition states required to assess the phase~\cite{OBrien2003}.

In our model, the unknown parameters to retrieve include the mode matrix $V$, the pair generation probability $\mu$, and the total system efficiencies $\eta_A$ and $\eta_B$ preceding detection at the control and target photon detectors, respectively. Obtained from independent measurements, and thus taken as fixed and known, are the dark count probabilities $d_A$ and $d_B$. All probabilities $\{\mu,d_A,d_B \}$ are specified for one resolving time $\tau$ ($\sim$1.5 ns). For the input photon state $|C_k T_l\rangle$ ($k,l\in\{0,1\}$) with detectors $A$ and $B$ set to respond to output modes $C_r$ and $T_s$ ($r,s\in\{0,1\}$), respectively, the probability of a coincidence between detectors $A$ and $B$ is
\begin{equation}
\label{e2-2}
p_{AB} = \mu \eta_A \eta_B \left|V_{C_r C_k} V_{T_s T_l} + V_{C_r T_l} V_{T_s C_k} \right|^2 + 2 p_A p_B.
\end{equation}
Here $p_A$ and $p_B$ are the marginal probabilities for clicks on $A$ or $B$, irrespective of clicks on the other, during a given time $\tau$. This formula thus contains both a correlated term (from photons of the same pair) and an accidental term. The latter, equal to $2p_A p_B$~\cite{Eckart1938}, represents the chance of simultaneous clicks in which at least one detector registers a dark count, or the photons come from different pairs (see Appendix C for details).


The marginal probabilities $p_A$ and $p_B$ can be found by summing the contributions from each possible number of photons $N$ being present in the monitored mode, sketched formally as, e.g., $p_A = \sum_N P(\mathrm{click}|N\;\mathrm{photons}) P(N\;\mathrm{photons})$. Writing out each term for $N=0,1,2$, and simplifying, we ultimately arrive at the probabilities for a click on either detector within a time $\tau$ (see Appendix C):
\begin{equation}
\begin{aligned}
p_A = \mu \eta_A \left( \left|V_{C_r C_k} \right|^2 + \left|V_{C_r T_l} \right|^2 \right) + d_A
\\ 
\label{e2-1}
p_B = \mu \eta_B \left( \left|V_{T_s C_k} \right|^2 + \left|V_{T_s T_l} \right|^2 \right) + d_B,
\end{aligned}
\end{equation}
valid under the assumptions $\mu,\eta_A,\eta_B,d_A,d_B \ll 1$---satisfied in our experiment. In words, a detector can click from either of the following: (i) a photon pair is generated ($\mu$), one of the photons is sent to the monitored frequency bin (through $V$), and the photon reaches the output and is successfully detected ($\eta_A,\eta_B$); or (ii) the detector fires spontaneously ($d_A,d_B$). Crucially, the singles probabilities [Eq.~(\ref{e2-1})] depend only on the moduli of the $V$-matrix elements, whereas the coincidences also depend on the relative phase [via the permanent term in Eq.~(\ref{e2-2})]. It is this complementary dependence which underpins our ability to extract the full complex matrix from experimental data.


Specifically, for a single preparation/measurement configuration we possess three numbers as data: clicks on $A$ ($N_A$), clicks on $B$ ($N_B$), and coincidences ($N_{AB}$). This gives us the multinomial likelihood for this specific input/output configuration ($|C_k T_l\rangle \rightarrow |C_r T_s\rangle$):
\begin{align}
\label{e2-3}
P\left(\mathcal{D}_{C_k T_l}^{C_r T_s}\Big|\beta\right) =  (p_A-p_{AB})^{N_A - N_{AB}}  (p_B-p_{AB})^{N_B-N_{AB}} \nonumber \\ \times p_{AB}^{N_{AB}} (1-p_A-p_B+p_{AB})^{M-N_A-N_B+N_{AB}},
\end{align}
where $\mathcal{D}_{C_k T_l}^{C_r T_s}=\{N_A,N_B,N_{AB}\}$ contains all data values for the specific configuration. We have also reexpressed the events to make them mutually exclusive: click on $A$ only, happening $N_A-N_{AB}$ times; click on $B$ only, occurring $N_B-N_{AB}$ times; coincidence between $A$ and $B$ ($N_{AB}$ times); and no clicks (all remaining frames). The symbol $\beta$ is shorthand for all model parameters ($\beta = \{V,\mu,\eta_A,\eta_B\}$), and $M$ equals the total number of $\tau$ frames considered in one counting period (${\sim4}\times 10^{11}$ in our tests). The complete likelihood comprises 16 factors in the manner of Eq.~(\ref{e2-3}) for all combinations of inputs and outputs. Admittedly, our model does not account for completely general quantum processes; that is, we do not search for solutions from the set of all possible two-qubit operations, nor do we incorporate errors into the prepared input states. Nevertheless, our total channel model---linear-optical multiport plus dark-count and multipair noise effects---is strongly justified from physical considerations and is validated \emph{ex post facto} by the agreement with experiment below.

In order for our estimates to best reflect the measured photon-counting data, we assume uniform prior distributions over the interval $(0,1)$ for the unknown parameters $\mu$, $\eta_A$, and $\eta_B$. For the complex matrix $V$, we express each element in terms of amplitude and phase: $V_{nn^\prime}=r_{nn^\prime} e^{i\phi_{nn^\prime}}$. Since an overall scaling factor on $V$ is indistinguishable from changes to $\eta_A$ and $\eta_B$ [Eqs.~(\ref{e2-2},\ref{e2-1})], for concreteness we fix the Hilbert-Schmidt norm $\Tr V^\dagger V = 1.6558$, to match the ideal $V$ matrix obtained from the numerical optimization [see Eq.~(\ref{theory})], thus constraining the sum of the squares of $r_{nn^\prime}$. Otherwise, we let the squared amplitudes vary uniformly over all possible values subject to this condition. Because phase shifts on each of the modes before and after the multiport $V$ are not physically significant, we are free to take some of the $\phi_{nn^\prime}$ as given as well~\cite{Rahimi2013}. For convenience, we fix $\{\phi_{C_0 C_0}, \phi_{C_1 C_1}, \phi_{C_1 T_0}, \phi_{C_1 T_1}, \phi_{T_0 C_1}, \phi_{T_1 C_1} \}$ to their theoretical predictions, thus leaving 10 phases to be retrieved via BME.

With the likelihood and prior formally defined, in principle we are done: we have the posterior probability distribution from Bayes' rule, which represents complete knowledge of the parameters given the observed data. However, practically speaking, computing integrals or, equivalently, sampling from this many-parameter multimodal distribution is a formidable challenge. It is here that the techniques of Markov chain Monte Carlo (MCMC) sampling offer a solution, which---with minimal input---enable Bayesian machine learning of complex models. In our case, we employ slice sampling, an MCMC algorithm designed to produce a sequence of samples whose stationary distribution converges to the posterior~\cite{Neal2003}.

\begin{figure}
\centering
\includegraphics[width=\columnwidth]{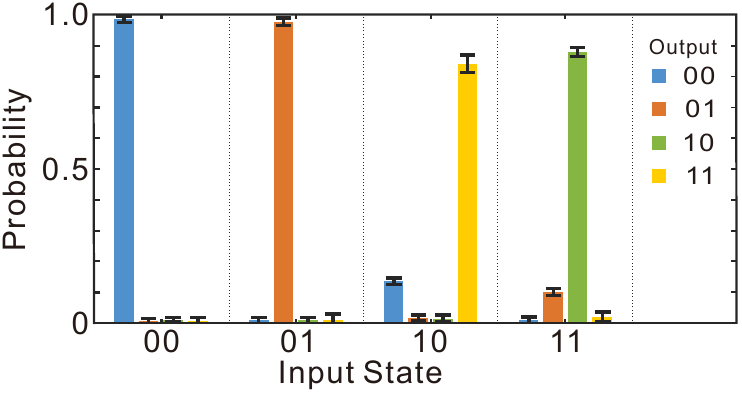}
\caption{Output state probabilities retrieved from BME, for each computational-basis input state.}
\label{fig4}
\end{figure}

Using the predicted matrix $V$ as an initial guess for the slice sampler, a procedure which we found important to speed up convergence given the large search space of 28 independent variables, we ultimately converge to the Bayesian fidelity estimate $\mathcal{F}_\mathrm{BME} = 0.91 \pm 0.01$, where $\mathcal{F}$ is defined according to Eq.~(\ref{e2}). Our truly quantum measurement does not reach the $>$0.99 classically inferred $\mathcal{F}_\mathrm{inf}$, which is a consequence of the relatively few coincidence counts ($<$100 in all cases) and additional noise from residual light. Nevertheless, the low uncertainty on $\mathcal{F}_\mathrm{BME}$ indicates high confidence in our BME model, especially in light of its ability retrieve the full complex fidelity with computational basis measurements. To see how  $\mathcal{F}_\mathrm{BME}$ translates into output state probabilities in the coincidence basis, we plot the Bayesian-estimated pathway probabilities in Fig.~\ref{fig4}, where the four outcomes for each input state are normalized to sum to unity. The average probability for obtaining the correct output is $0.92\pm 0.01$, computed by taking the mean of the four peaks in Fig.~\ref{fig4}. (See Appendix D for details on all retrieved parameters, including the mode matrix $V$.)

Moving forward toward implementation in a full QIP system, it will be important to mitigate the known source and detector noise contributions (i.e., dark counts and multipair emission). Currently, the high loss of our cascaded EOM/PS system will make this a challenge, but there is promise for significantly lower loss in an on-chip system. For example, process design kits from photonics foundries~\cite{Aim2018} suggest that the loss through a pulse shaper channel can be less than 1 dB, while a recent demonstration~\cite{Kieninger2018} indicates foundry-compatible EOMs with losses on the order of 1--2 dB.

In conclusion, we have realized the first entangling gate on frequency-bin qubits. We confirm high-fidelity operation of the CNOT with two forms of characterization: coherent-state-based matrix retrieval and photon pair measurements in the computational basis. The classically inferred fidelity of $\mathcal{F}_\mathrm{inf} = 0.995 \pm 0.001$ and Bayesian estimate $\mathcal{F}_\mathrm{BME}=0.91\pm 0.01$ both demonstrate high performance in our system. As the sole realization of a two-photon entangling gate in frequency---and only the second CNOT in the entire field of time-frequency quantum information~\cite{Humphreys2013}---our gate significantly expands the potential of single-spatial-mode, fiber-optic-based QIP. More generally, our Bayesian characterization approach provides further evidence of the potential of machine learning in analyzing quantum systems, particularly for extracting information within measurements which traditional methods overlook.


\textit{Acknowledgments.---}We thank N. Lingaraju for discussions. This work was performed in part at Oak Ridge National Laboratory, operated by UT-Battelle for the U.S. Department of Energy under contract no. DE-AC05-00OR22725. Funding was provided by ORNL's Laboratory Directed Research and Development Program.



\appendix

\section{Gate design.}

The optimization approach for designing quantum frequency gates using a series of EOMs/PSs was first proposed in Ref.~\cite{Lukens2017}, and adopted to experimentally demonstrate a single-photon gate in Ref.~\cite{Lu2018a}. In this work, we follow the same procedures, utilizing the MATLAB Optimization Toolbox to search for an optimal set of phases for a particular EOM/PS sequence, constraining fidelity $\mathcal{F}\geq0.9999$ and maximizing the success probability $\mathcal{P}$ for the two-photon state transformation matrix $W$. Compared to single-qubit gates, where only one frequency scale appears (the spacing between the two computational bins), two-qubit gates provide a much richer parameter space; namely, the placement of the four computational modes relative to each other can have a profound impact on the EOM/PS complexity needed to realize a specific operation. We have performed a thorough---though non-exhaustive---search over these possible mode placement combinations in each round of optimization. In general, we are guided by the intuition to spectrally isolate control mode 0 ($C_0$) while packing control mode 1 ($C_1$) close to both target modes. 

\begin{figure}[!b]
\includegraphics[width=\columnwidth]{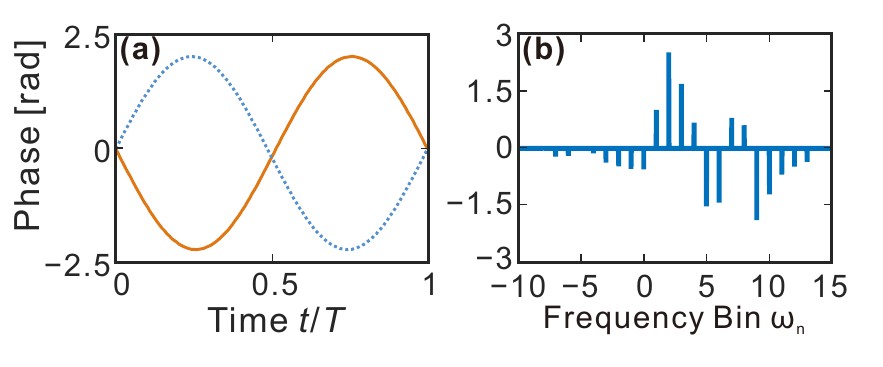}
\caption{Numerical solutions for the time-frequency phases required to implement coincidence-basis CNOT gate. (a) Temporal phase modulation applied to the first EOM (solid red) and second EOM (dotted blue), plotted over one period. (b) Phases applied to each frequency mode by the pulse shaper, where modes 0 and 6 denote the control bins $\{C_0,C_1\}$, and modes 7 and 8 represent the target bins $\{T_0,T_1\}$.}
\label{figE1} 
\end{figure}

For reference, the ideal CNOT matrix is
\begin{equation}
\label{eCNOT}
U_\mathrm{ideal} = \begin{bmatrix}
1 & 0 & 0 & 0 \\
0 & 1 & 0 & 0 \\
0 & 0 & 0 & 1 \\
0 & 0 & 1 & 0
\end{bmatrix},
\end{equation}
against which we compare the numerically obtained two-photon matrix $W$ (a function of $V$) via Eq.~(\ref{e2}). The optimal solution we found for the CNOT gate using a 2EOM/1PS circuit is presented in Fig.~\ref{figE1}, with $\mathcal{F} =0.9999$, $\mathcal{P} =0.0445$, and modes $\{C_0,C_1,T_0,T_1\}$ at frequency bins $\{0,6,7,8\}$, respectively. The temporal phase modulation on both EOMs are simply $\pi$-phase-shifted sinewaves, and combined with the spectral phase modulation imparted by the PS, the corresponding mode transformation matrix $V$ is numerically calculated as:
\begin{widetext}
\begin{equation}
\label{theory}
V = \left[r_{nn^\prime}\measuredangle\phi_{nn^\prime}\right] = \begin{bmatrix}
{0.4407\measuredangle{-2.5976}} & 0.0022\measuredangle0.2103 & 0.0026\measuredangle1.2938 &  0.0010\measuredangle{-2.0353} \\
0.0022\measuredangle0.2104 & {0.4343\measuredangle{-2.6045}} & {0.4596\measuredangle{-1.5754}} & {0.4549\measuredangle1.5710} \\
0.0026\measuredangle1.2939 & {0.4596\measuredangle{-1.5754}} & {0.4830\measuredangle2.5973} & 0.0030\measuredangle{-2.8778} \\ 
0.0010\measuredangle{-2.0352} & {0.4549\measuredangle1.5710} & 0.0030\measuredangle{-2.8779} & {0.4783\measuredangle2.5979}
\end{bmatrix},
\end{equation}
\end{widetext}
using the phasor shorthand $r_{nn^\prime}\measuredangle\phi_{nn^\prime} \equiv r_{nn^\prime}e^{i\phi_{nn^\prime}}$. This provides a reference to compare the experimental mode transformations below, obtained either by coherent state characterization or BME.

\section{Coherent state measurements.}
To investigate the performance of a linear-optical multiport, Ref.~\cite{Rahimi2013} provides an efficient characterization method utilizing only coherent states as sources and power measurements at the output. We follow similar procedures (adopted in Ref.~\cite{Lu2018a} for single-qubit frequency gates) by probing our frequency multiport with an electro-optic frequency comb, and measuring the output spectrum for different input frequency superpositions. We first send a continuous-wave laser with center frequency $\Omega_{01} = 2\pi \times 193.550$ THz [see Fig.~\ref{fig2}(a)] into an additional EOM modulated at 25 GHz to create $\sim$10 comb lines, and we utilize a subsequent pulse shaper to prepare specific input states. To obtain the modulus of every matrix element in the four columns of $V$, each time we send in only one input mode from the set $\{C_0,C_1,T_0,T_1\}$ and measure the spectrum at the output of the gate, collecting all the output modes with power levels within 60~dB of the maximum. This allows us to retrieve the amplitudes $r_{nn^\prime}$. Then by sending in two lines and scanning their relative input phase, we can map out the $V$-matrix phases $\phi_{nn^\prime}$, where we compute all unknown values relative to phase values we are free by physical considerations to define \emph{a priori}~\cite{Rahimi2013}. We perform five identical measurements of $V$ in order to estimate uncertainty; following are the resulting amplitudes and phases, with each number averaged individually over the five successive, independent matrix acquisitions:
\begin{widetext}
\begin{equation}
\label{CSamp}
\left[r_{nn^\prime}\right] = \begin{bmatrix}
0.428\pm0.008 & 0.0030\pm0.0003 & 0.0027\pm0.0001 &  0.0017\pm0.0001 \\
0.0031\pm0.0001 & 0.427\pm0.001 & 0.451\pm0.002 & 0.451\pm0.002 \\
0.0028\pm0.0002 & 0.465\pm0.005 & 0.478\pm0.003 & 0.041\pm0.003 \\
0.0018\pm0.0003 & 0.458\pm0.002 & 0.036\pm0.004 & 0.499\pm0.006
\end{bmatrix}
\end{equation}
\begin{equation}
\label{CSphase}
\left[\phi_{nn^\prime}\right] = \begin{bmatrix}
-2.5976\pm0 & $\ldots$ & $\ldots$ & $\ldots$ \\
$\ldots$ & -2.6045\pm0 & -1.5754\pm0 & 1.5710\pm0 \\
$\ldots$ & -1.5754\pm0 & 2.621\pm0.002 & -2.89\pm0.05 \\
$\ldots$ & 1.5710\pm0 & -2.7\pm0.1 & 2.631\pm0.006
\end{bmatrix}.
\end{equation}
\end{widetext}
The phase values with $\pm0$ uncertainty are those we could fix to the theoretical prediction [Eq.~(\ref{theory})], found by the optimizer to yield high CNOT fidelity. Because the coupling between mode $C_0$ and $\{C_1,T_0,T_1\}$ is too weak, we could not extract meaningful phase estimates of the elements delineated by ``$\ldots$'' in the $\phi_{nn^\prime}$ matrix. However, we have confirmed that setting these phases to any set of random values impacts our calculation of the fidelity at only the fifth decimal place, so that it has no influence on our computed $\mathcal{F}_\mathrm{inf} = 0.995\pm0.001$. From the retrieved amplitudes and phases, we find uncertainties for the eight large elements ($r_{nn^\prime}>0.4$) at the third significant digit, an indication of the high precision possible with this high-flux characterization method.

\section{Parameter model.} 
In order to make use of the observed data to estimate the key parameters of our quantum gate, we first derive a realistic model connecting the underlying gate operation to photon counts, encapsulated in a likelihood function $P(\bm{\mathcal{D}}|\beta)$, for the model parameters $\beta$ given data $\bm{\mathcal{D}}$ (proportional to the conditional probability of $\bm{\mathcal{D}}$ given $\beta$). In our case, the set $\beta$ contains not only the mode transformation matrix $V$, but also the pair generation probability $\mu$ and the system efficiencies $\eta_A$ and $\eta_B$.

Initially, we focus on how the input quantum state propagates through the multiport---for the moment neglecting loss, which will be incorporated later. The total optical network (defined over countably infinite frequency bins) maps inputs $\hat{a}_n$ to outputs $\hat{b}_n$ according to
\begin{equation}
\label{e1-x}
\hat{b}_n = \sum_{n^\prime=-\infty}^{\infty} V_{nn^\prime} \hat{a}_{n^\prime},
\end{equation}
with $V$ unitary when considered over all modes. For a particular counting experiment, we take the prepared input state as
\begin{equation}
\label{e2-y}
|\Psi\rangle = |1_u 1_v\rangle = \hat{a}_u^\dagger \hat{a}_v^\dagger |\mathrm{vac}\rangle,
\end{equation}
where $u\neq v$. Specifying such a state assumes that: (i)  contributions from other frequency-bin pairs can be neglected, justified experimentally by the BFC shaper's ability to suppress adjacent frequency bins by $>$40~dB; and (ii) higher-order pair generation (e.g., four, six, eight, etc., photon terms) need not be included explicitly. Incidentally, the ansatz we incorporate for accidental coincidences [see Eq.~(\ref{e12}) below] ends up capturing the main effects of multiple photon pairs on our data in a simpler fashion.

We define $p_\mu (1_m 1_n)$ as the probability for one photon to be found in mode $m$ and the other in mode $n$ at the output (again assuming no loss). This is given by
\begin{equation}
\label{e3}
p_\mu (1_m 1_n) = \left|V_{mu} V_{nv} + V_{mv} V_{nu} \right|^2 \;\;\;\; (n\neq m).
\end{equation}
When $n=m$ (two photons in the same mode), the probability is
\begin{equation}
\label{e4}
p_\mu (2_m) = 2 \left| V_{mu}V_{mv} \right|^2,
\end{equation}
with the factor of two a consequence of boson statistics. From these results, we can also compute the marginal probability for one-photon occupancy in a particular mode,
\begin{equation}
\begin{aligned}
\label{e5}
p_\mu (1_m) 
& = \sum_{\substack{ n=-\infty \\ n\neq m}}^{\infty} \left|V_{mu} V_{nv} + V_{mv} V_{nu} \right|^2  \\
& = \sum_{n=-\infty}^{\infty} \Big( \left|V_{mu} V_{nv} + V_{mv} V_{nu} \right|^2 \Big) - 4|V_{mu}V_{mv}|^2  \\
& = |V_{mu}|^2 + |V_{mv}|^2 - 4|V_{mu} V_{mv}|^2,
\end{aligned}
\end{equation}
with the last line following from the unitarity of $V$ and the fact that $u\neq v$ in our input state.

We then map these fundamental ``per-pair'' probabilities to expected detection rates. For accounting purposes, we define all detection probabilities within a specific temporal frame $\tau$, the time within which clicks on detector $A$ ($t_A$) and $B$ ($t_B$) are deemed coincident: $|t_A-t_B| < \tau$. Our stationary (continuous-wave pumped) source ensures that all such probabilities are equal in every length-$\tau$ time bin. With $\mu$ defined as the pair generation probability within such a frame, the marginal probabilities for single-detector clicks are
\begin{equation}
\begin{aligned}
\label{e8}
p_A & = \mu\left[ \eta_A + (1-\eta_A) \eta_A \right] p_\mu(2_m) + \mu\eta_A p_\mu (1_m) + d_A  \\
p_B & = \mu\left[ \eta_B + (1-\eta_B) \eta_B \right] p_\mu(2_n) + \mu\eta_B p_\mu (1_n) + d_B
\end{aligned}
\end{equation}
for detector $A$ monitoring frequency bin $m$ and $B$ frequency bin $n$. The probabilities $d_A$ and $d_B$ represent the dark (or more generally, background) count probabilities; we measure these independently and take them as fixed at $d_A=9.60\times 10^{-7}$ and $d_B=7.77\times 10^{-7}$. The efficiencies $\eta_A$ and $\eta_B$ include all loss effects through the system, from generation in the crystal to photon detection; we assume them to be mode-independent---validated by the relatively small bandwidth comprising all modes of interest ($\sim$500~GHz)---yet they can vary by the different relative efficiencies of our superconducting nanowire detectors. And while spectral filtering \emph{per se} does not modify these general considerations, simple theory suggests that the Lorentzian linewidth profile of the etalon will introduce an effective transmission smaller than its peak value---we believe this contributes to lower overall $\eta_{A}$ and $\eta_B$ retrieved in BME. Next we make use of the fact that the system efficiencies $\eta_A, \eta_B \ll 1$. Plugging in Eqs.~(\ref{e4}) and (\ref{e5}), we obtain
\begin{equation}
\begin{aligned}
\label{e9}
p_A = \mu\eta_A \Big( |V_{mu}|^2 + |V_{mv}|^2 \Big) + d_A \\
p_B = \mu\eta_B \Big( |V_{nu}|^2 + |V_{nv}|^2 \Big) + d_B.
\end{aligned}
\end{equation}
The simple addition of pair and dark-count contributions is justified in our case by their small values ($\sim$10$^{-6}$), so that there is no concern for $p_{A}$ or $p_B$ approaching or exceeding 1 in the numerical analysis below.

To establish the probability for a coincidence between detectors $A$ and $B$ in our model, we make a sharp distinction between two types of events: (i) correlated coincidences, deriving from two photons of the same pair; and (ii) accidental coincidences, in which two random clicks (from at least one dark count, or photons from two different pairs) overlap within the resolving time $\tau$. We note that, in principle, such a distinction is not necessary: it should be possible to derive a completely \emph{ab initio} model for coincidences, with an input density matrix including higher-order pair generation effects, and positive-operator valued measures (POVMs) incorporating dark count noise. However, our approach proves much simpler, requiring fewer parameters while still satisfying conceptual demands.

For event (i), the click probability follows from multiplying the per-pair probability $p_\mu(1_m 1_n)$ by $\mu \eta_A \eta_B$, so that $p_{AB}^{(\mathrm{i})} = \mu \eta_A \eta_B \left|V_{mu}V_{nv}+V_{mv}V_{nu} \right|^2$, which assumes that $\tau$ is sufficiently large to integrate over the full two-photon correlation time. Regarding event (ii), in general the rate of accidental coincidences between two independent detectors is given by a product of the rates of the two detectors individually: $R_{AB}^{(\mathrm{ii})} = 2\tau R_A R_B$~\cite{Eckart1938}, where the factor of two follows from the fact that---under our definition of $\tau$---all events such that $(t_A-t_B) \in (-\tau,\tau)$ register as coincidences. Making the connection $p_{j}=\tau R_{j}$ then allows us to write $p_{AB}^{(\mathrm{ii})} = 2p_A p_B$, so that the total coincidence probability becomes
\begin{equation}
\begin{aligned}
\label{e12}
p_{AB} & = p_{AB}^{(\mathrm{i})} + p_{AB}^{(\mathrm{ii})} \\
 & = \mu \eta_A \eta_B  \Big|V_{mu}V_{nv}+V_{mv}V_{nu} \Big|^2 + 2 p_A p_B,
\end{aligned}
\end{equation}
with $p_A$ and $p_B$ defined as in Eq.~(\ref{e9}). Expanding $2p_A p_B$, the expected noise sources appear naturally: a $\mu^2$ term reflects clicks from two different pairs, while $\mu d_A$ and $\mu d_B$ terms give coincidences from a photon and dark count. In this way, we can recover noise effects otherwise absent in the physical model, via what can be called an ``accidentals correction'' term $2p_A p_B$. Finally, we emphasize that the accuracy of Eq.~(\ref{e12}) relies again on the relative order of magnitudes of the probabilities involved: $p_{AB}^{(\mathrm{i})} \sim 10^{-10}$, so that the differences between alternative forms one could conceivably argue for---such as $p_B \rightarrow p_B - p_{AB}^{(\mathrm{i})}$, to help ensure that singles counts from correlated coincidences do not also count toward accidental probabilities---become numerically inconsequential.

Finally, with these probabilities established, we can write the likelihood using a multinomial distribution for all event types. Over the course of a single measurement of duration $T$, we experience $M=T/\tau$ total frames, in which we can register one of the four mutually exclusive outcomes: click on $A$ only, click on $B$ only, coincidence, or no clicks. The likelihood for the specific input/output mode configuration (defined by the mode numbers $uv \rightarrow mn$) is
\begin{equation}
\begin{aligned}
\label{e14}
P\left(\mathcal{D}_{uv}^{mn}|\beta\right) & = (p_A - p_{AB})^{N_A-N_{AB}} (p_B - p_{AB})^{N_B-N_{AB}} \\
& \times p_{AB}^{N_{AB}} (1 - p_A - p_B + p_{AB})^{M-N_A-N_B + N_{AB}},
\end{aligned}
\end{equation}
where we emphasize that both the dataset $\mathcal{D}_{uv}^{mn}=\{N_A,N_B,N_{AB}\}$ and probabilities $\{p_A,p_B,p_{AB}\}$ themselves depend on the mode configuration $uvmn$. The total likelihood follows by multiplying out all 16 individual combinations
\begin{equation}
\label{e15}
P\left(\bm{\mathcal{D}}|\beta\right) = \prod_{\substack{u,m\in\{C_0,C_1\} \\ v,n \in \{T_0, T_1\}}} P\left(\mathcal{D}_{uv}^{mn}|\beta\right),
\end{equation}
where the modes $\{C_0,C_1,T_0,T_1\}$ are as defined in the main text. (We also neglect unimportant scaling factors which do not depend on the parameters $\beta$.) This likelihood forms the basis for estimating the parameters $\beta=\{V,\mu,\eta_A,\eta_B\}$ from the dataset $\bm{\mathcal{D}}=  \bigcup \mathcal{D}_{uv}^{mn}$.

\section{Bayesian machine learning.}
To estimate these values along with their uncertainties, we make use of Bayes' rule for the posterior probability distribution
\begin{equation}
\label{e16}
P(\beta|\bm{\mathcal{D}}) =  \frac{1}{\mathcal{Z}} P\left(\bm{\mathcal{D}}|\beta\right) P(\beta),
\end{equation}
with $\mathcal{Z}=\int d\beta P\left(\bm{\mathcal{D}}|\beta\right) P(\beta)$ the (undetermined) normalizing factor. $P(\beta)$ represents the prior probability distribution for the parameters. We take $P(\beta)$ as uniform over $(0,1)$ for each of $\mu$, $\eta_A$, and $\eta_B$; uniform over $(0,2\pi)$ for all phases $\phi_{nn^\prime} = \arg V_{nn^\prime}$ which are not taken as fixed $\{\phi_{C_0 C_0}, \phi_{C_1 C_1}, \phi_{C_1 T_0}, \phi_{C_1 T_1}, \phi_{T_0 C_1}, \phi_{T_1 C_1} \}$; and uniform for all squared moduli $r_{nn^\prime}^2$ subject to the constraint $\sum_{nn^\prime} r_{nn^\prime}^2=1.6558$ from Eq.~(\ref{theory}). This uninformative prior allows the estimates to be fully determined by the counting data itself.

Due to the complexity of integrating Eq.~(\ref{e16}) over our parameter space, we employ slice sampling~\cite{Neal2003} and retrieve 4096 samples of all 28 parameters from the unnormalized  $P\left(\bm{\mathcal{D}}|\beta\right) P(\beta)$. We use best guesses of all parameters as the starting point to enable convergence, invoking a burn-in period and thinning until stationarity is achieved. At each sample of $\beta$, we can compute any quantity of interest, and use the statistics over all samples to produce the mean and standard deviation. Specifically, we find
\begin{equation}
\begin{aligned}
\mu & = 0.024 \pm 0.002 \\
\eta_A & = (3.5 \pm 0.3) \times 10^{-4} \\
\eta_B & = (4.7 \pm 0.3) \times 10^{-4} \\
\mathcal{F}_\mathrm{BME} & = 0.91 \pm 0.01.
\end{aligned}
\end{equation}
The retrieved pathway efficiencies are smaller by $\sim$9~dB compared to our insertion loss alone, which we estimate to be $\sim$25 dB from generation to detection. While we have fully characterized the insertion loss of the gate components themselves (12.9~dB~\cite{Lu2018a}), uncertainties remain in the state preparation and measurement components, such as the breakdown of loss inside the fiber-pigtailed photon source, as well as questions of how strongly the spectrally varying transmission of the etalon reduces its effective transmission from its peak value. Otherwise, the retrieved $\mu$ and fidelity match predictions. Even though $\mathcal{F}_\mathrm{BME}$ is smaller and has higher uncertainty than the classically inferred $\mathcal{F}_\mathrm{inf}$, the fact it still exceeds $90\%$ with fairly sparse measurements is strong confirmation of excellent performance, particularly in light of the uninformative prior, which permits high fidelity only based on the strength of the observed data.

We also compute the mean and standard deviation for all elements of the retrieved transformation $V$, for both the magnitude and phase:
\begin{widetext}
\begin{equation}
\left[r_{nn^\prime}\right]  = \begin{bmatrix}
0.452\pm0.005 & 0.124\pm0.009 & 0.06\pm0.01 &  0.02\pm0.02 \\
0.06\pm0.03 & 0.465\pm0.008 & 0.475\pm0.006 & 0.411\pm0.006 \\
0.04\pm0.01 & 0.463\pm0.005 & 0.470\pm0.005 & 0.03\pm0.01 \\
0.028\pm0.009 & 0.455\pm0.005 & 0.02\pm0.01 & 0.413\pm0.005
\end{bmatrix}
\end{equation}
\begin{equation}
\left[\phi_{nn^\prime}\right]  = \begin{bmatrix}
-2.5976\pm0 & -2.8\pm0.2 & 1.3\pm0.1 & -2.01\pm0.09 \\
0.30\pm0.09 & -2.6045\pm0 & -1.5754\pm0 & 1.5710\pm0 \\
1.35\pm0.09 & -1.5754\pm0 & 2.6\pm0.1 & 0.7\pm0.2 \\
-2.0\pm0.1 & 1.5710\pm0 & 0.3\pm0.1 & 2.5\pm0.1
\end{bmatrix}.
\end{equation}
\end{widetext}
As before, the phases with uncertainties $\pm0$ are those fixed prior to parameter retrieval. Comparing this result to the design [Eq.~(\ref{theory})] and coherent-state-retrieved matrix [Eqs.~(\ref{CSamp},\ref{CSphase})], the most significant mismatch occurs for the element in row 1, column 2 (the coupling from mode $C_1$ to $C_0$). At 0.124, this value is significantly larger than designed, and contributes to the higher error for the cases $|C_1T_0\rangle \rightarrow |C_0T_0\rangle$ and $|C_1T_1\rangle \rightarrow |C_0T_1\rangle$ in Fig.~(\ref{fig4}). While the source of this error is still uncertain, experimentally we did observe extraneous counts on detector $A$ during these integration times, beyond the theoretical prediction. Bayesian retrieval succeeds in finding matrix elements to account for this observation, as intended.

\newpage

%

\end{document}